\documentclass[final,5p,twocolumn]{elsarticle}
\usepackage{amsmath}
\journal{Physica D}
\usepackage{graphicx}
\usepackage{epstopdf}
\usepackage{natbib}

\begin{document}
\begin{frontmatter}

\title{Axisymmetric pulse train solutions in narrow-gap spherical Couette flow}
\author{Adam Child\corref{cor1}}
\author{Rainer Hollerbach}
\author{Evy Kersal\'e}
\address{Department of Applied Mathematics, University of Leeds, Leeds, LS2 9JT, UK}

\begin{abstract}
We numerically compute the flow induced in a spherical shell by fixing the outer sphere and rotating the inner one. The aspect ratio $\epsilon=(r_o-r_i)/r_i$ is set at 0.04 and 0.02, and in each case the Reynolds number measuring the inner sphere's rotation rate is increased to $\sim$10\% beyond the first bifurcation from the basic state flow. For $\epsilon =0.04$ the initial bifurcations are the same as in previous numerical work at $\epsilon=0.154$, and result in steady one- and two-vortex states. Further bifurcations yield travelling wave solutions similar to previous analytic results valid in the $\epsilon\to0$ limit. For $\epsilon=0.02$ the steady one-vortex state no longer exists, and the first bifurcation is directly to these travelling wave solutions, consisting of pulse trains of Taylor vortices travelling toward the equator from both hemispheres, and annihilating there in distinct phase-slip events. We explore these time-dependent solutions in detail, and find that they can be both equatorially symmetric and asymmetric, as well as periodic or quasi-periodic in time.
\end{abstract}

\begin{keyword}
Spherical Couette flow \sep Hydrodynamic stability \sep Pattern formation
\end{keyword}
\end{frontmatter}

\section{Introduction}
How and why nonlinear systems form particular patterns is of enormous interest in a broad variety of applications \citep{Cross1993,Hoyle2006,Cross2009}. A recurring theme is the development of small-scale patterns on a background that varies on large scales. One fluid dynamical system in which this arises naturally is spherical Couette flow -- the flow between differentially rotating concentric spheres -- in the limit of a very narrow gap. The expected
scale for instabilities is then the gap width $r_o-r_i$, whereas the scale on which the background varies is the radius $r_i$, where $r_i$ and $r_o$ are the radii of the inner and outer spheres, respectively. If the aspect ratio
\begin{displaymath}
 \epsilon = (r_o-r_i)/r_i
\end{displaymath}
is sufficiently small, the scale separation between instabilities and background becomes arbitrarily large. Previous work has traditionally been classified into three regimes: wide gap $\epsilon>0.24$ \citep{Munson1975,Egbers1995,Hollerbach2006}, medium gap $0.12<\epsilon<0.24$ \citep{buhler1990,mamun1995,nakabayashi2005}, and narrow gap $\epsilon<0.12$ \citep{wimmer1976,wimmer1981,nakabayashi2002}. We will here concentrate on the very narrow gap regime $\epsilon\le0.04$, and demonstrate that differences in the first few bifurcations can occur even in this limited regime.

Wimmer \citep{wimmer1976,wimmer1981} performed experiments for gaps as thin as $\epsilon=0.0063$, and found that roughly circular Taylor vortices occurred in the equatorial regions. The ratio of length scales is thus indeed $\epsilon$, as expected. Asymptotic solutions \citep{walton1978,hocking1981} verified that the instability should
be localised to the equatorial region, and should occur for Reynolds numbers somewhat larger than the corresponding values in cylindrical geometry. This difference is due to the large scale variation (curvature) of the spherical geometry, and the resulting meridional Ekman circulation cells. This results in phase mixing, which serves to enhance dissipation and is thus a stabilising influence. Via utilisation of a WKBJ method, accurate asymptotic values
were first given by \citep{soward1983}. 

\citet{bartels1982} performed very early numerical simulations in the narrow gap limit, as low as $\epsilon=0.0256$, and obtained travelling wave solutions in which Taylor vortices propagate toward the equator. However, computational limitations in 1982 allowed only one single Reynolds number to be computed at the smallest $\epsilon=0.0256$. Nevertheless, these pioneering calculations form part of our motivation for re-examining the same problem. The second line of research motivating our work is a series of asymptotic analyses by Soward, Bassom, and co-workers, most recently summarized in \citep{soward2016}.

Specifically, \citep{harris2000,harris2003} examined the case where the spheres almost corotate ($\delta=(\Omega_i-\Omega_o)/\Omega_i\sim\epsilon^{1/2}$, where $\Omega_i$ and $\Omega_o$ are the angular velocities of the inner and outer spheres), also obtaining travelling wave solutions. \citet{bassom2004}, influenced by work on thermal convection \cite{ewen1994phase,ewen1994wave}, conclude that such solutions take the form of pulse trains. They developed a rigorous analytic theory in the $\epsilon\to0$ limit, for which it is shown that pulse train solutions exist when $\epsilon^{1/2}<\delta<1$, encompassing cases from almost co-rotation to a stationary outer sphere. Physically it was shown that, for $\epsilon\to0$, vortices exist in a region localised around the equator, under some wave-envelope with an amplitude proportional to the distance from the equator. There exists a dislocation halfway between pulse centres, and over long time scales these centres drift poleward, leading to unstable solutions.  \citet{blockley2007} expanded upon this to give a number of additional symmetry classes for the pulse train solutions, showing both periodic and chaotic behaviour, and reached the conclusion that such solutions are similar to the 3-D spiral vortices of \citep{nakabayashi1983transition,dumas1994,yuan2004numerical}. Indeed, the azimuthal evolution of the spiral vortices is similar to the time evolution of axisymmetric flow.

In this paper, we explore axisymmetric pulse train solutions for $\epsilon=0.04$ and 0.02, and map out the bifurcation sequences that arise. We show that $\epsilon=0.04$ still has some aspects in common with much wider gaps ($\epsilon=0.154$), but for $\epsilon=0.02$ the first bifurcation is directly to a pulse train solution similar to the asymptotics discussed above. Finally, we link the structure of the wave-envelope, under which vortices oscillate, to the bifurcation sequence and phase-slips present in the flow.

\section{Equations}
We consider axisymmetric spherical Couette flow, with radii $r_i$ and $r_o$ and angular velocities $\Omega_i$ and $\Omega_o$ at the inner and outer spheres, respectively. The nondimensional Navier-Stokes equation is
\begin{equation}
 \frac{\partial {\bf{U}}}{\partial t} - \mathrm{Re}^{-1}\nabla^2{\bf{U}} = -\nabla p - \textbf{U}\cdot\nabla\textbf{U},
\end{equation}
together with the incompressibility condition $\nabla\cdot\textbf{U}=0$. Length and time have been nondimensionalised using the inner radius $r_i$, and the rotational period $\Omega_i^{-1}$, respectively. The Reynolds number is thus defined as
\begin{equation}
 \mathrm{Re} = {r_i^2\Omega_i}/{\nu}.
\end{equation}
The boundary conditions are no-slip, $\textbf{U} = \Omega_{i,o}\,r_{i,o}\sin\theta\,{\bf\hat e}_\phi$ at the inner and outer spheres. 

Since we only consider axisymmetric flow, it proves advantageous to decompose the velocity field into toroidal and poloidal components,
\begin{equation}
 {\bf{U}} = v\,{\bf{e}}_\phi + \nabla\times(\psi\,{\bf{e}}_\phi),
\end{equation}
such that the incompressibility condition is automatically fulfilled. The quantities $v$ and $\psi$ are further decomposed into Chebyshev polynomials in $r$, and associated Legendre functions in $\theta$. Relatively little structure develops in the radial direction, so $\sim$20 Chebyshev polynomials were sufficient for all solutions considered here. In latitude though we require up to 2000 Legendre functions, since we are specifically interested in the small $\epsilon$ regime where the contrast between large and small scales is substantial. At such large $\theta$ resolutions, the Legendre transforms used by \citep{hollerbach2000} turned out to be too inefficient. A new code was therefore developed based on much the same general method, but using the highly optimised `SHTNS' Legendre transforms of \citet{schaeffer2013}. The resulting code was also benchmarked against results from \citep{hollerbach1998}, with agreement to within a fraction of one percent.

For this work, we focus solely on the case where only the inner sphere is rotating, $\Omega_0=0$, which corresponds to the $\delta=1$ results discussed by \citep{bassom2004} and \citep{blockley2007}. It should be noted that many of their solution branches are qualitatively similar regardless of whether the outer sphere is stationary or co-rotating. We focus on the two values $\epsilon=0.04$ and 0.02, and in each case increase $\mathrm{Re}$ to $\sim$10\% above the first bifurcation. Based on the previous asymptotic work \citep{harris2003,bassom2004}, we expect the critical Reynolds numbers to be around 5400 for $\epsilon=0.04$, and 15100 for $\epsilon=0.02$, after translating from their Taylor number notation to our $\mathrm{Re}$.

\section{Results}

\begin{figure}[ht]
 \centering
 \includegraphics[width=\columnwidth]{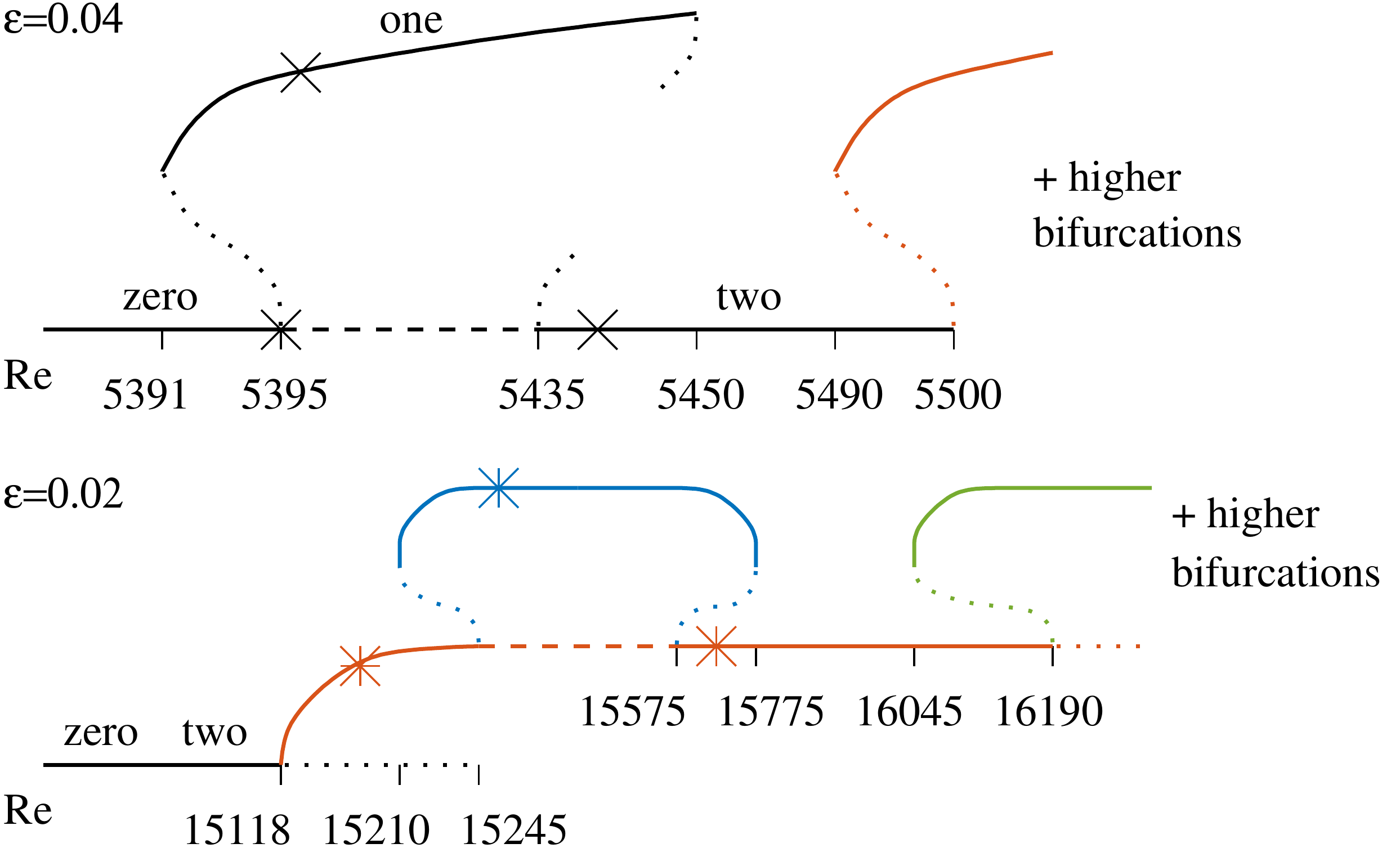}
 \caption{Schematic bifurcation diagrams for $\epsilon=0.04$ (top) and $\epsilon=0.02$ (bottom). Solid lines indicate stable solutions, dashed lines indicate verified unstable solutions, and dotted lines indicate presumed unstable solutions. The colours represent different symmetries and time dependencies in the solutions: black is symmetric steady state, red is symmetric periodic, blue is asymmetric periodic, and green is asymmetric quasiperiodic. Crosses correspond to solutions shown in figure \ref{fig:contours}; asterisks denote solutions in figure \ref{fig:hov_per}.}
 \label{fig:bifurc}
\end{figure}

\begin{figure}[ht]
 \centering
 \includegraphics[width=0.75\columnwidth]{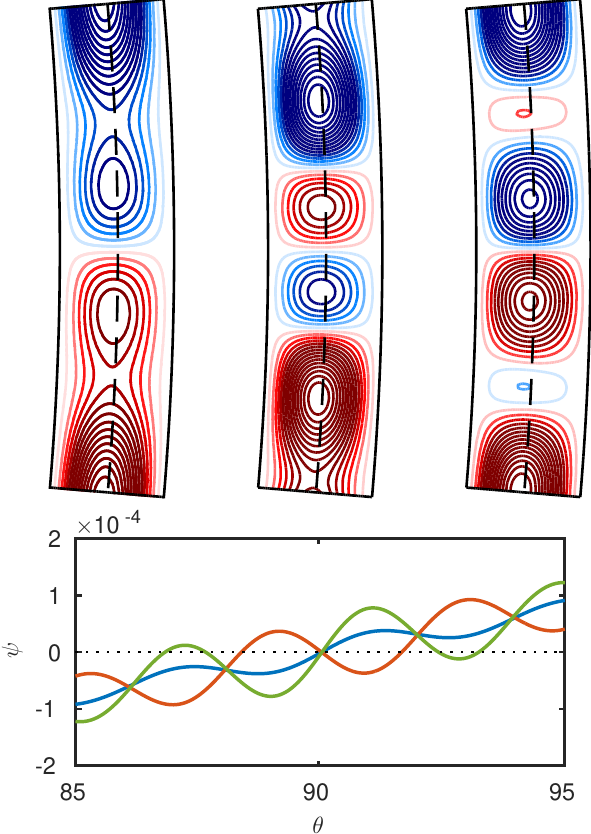}
 \caption{The top row shows contours of meridional circulation $\psi$ for zero-, one- and  two-vortex states at $\mathrm{Re}=5395$, 5400 and 5440, respectively, and all three at  $\epsilon=0.04$. The bottom row shows $\psi$ as a function of $\theta$ only (varying  $\pm5^\circ$ from the equator), at the particular value of $r$ indicated by the dotted lines  in the top row. The three curves are color coded according to: blue, red, green corresponds  to zero, one, two states, respectively.}
 \label{fig:contours}
\end{figure}

Figure \ref{fig:bifurc} shows schematic bifurcation diagrams for $\epsilon=0.04$ and 0.02. In both cases the basic state before any bifurcations occur is the so-called zero-vortex state, having only one extremely elongated Ekman circulation cell in each hemisphere, but no Taylor vortices. Turning to $\epsilon=0.04$ first, we see that this basic state becomes unstable via a subcritical pitchfork bifurcation at $\mathrm{Re}=5395$. The intermediate stages of the subsequent evolution are equatorially asymmetric, but the system eventually equilibrates to an equatorially symmetric state again, the so-called one-vortex state, having one Taylor vortex in each hemisphere (so two vortices in total). If $\mathrm{Re}$ is subsequently reduced again, this one-vortex branch exhibits a slight amount of hysteresis, existing back down to a turning point bifurcation at $\mathrm{Re}=5391$. If instead $\mathrm{Re}$ is increased, the one-vortex state remains stable up to $\mathrm{Re}\approx5450$, where it also becomes unstable to a subcritical
pitchfork bifurcation. The subsequent transients again involve equatorial asymmetry, but the final equilibrated solution is again equatorially symmetric, the two-vortex state, with two Taylor vortices in each hemisphere. This transition between the one state and the two state also involves a slight degree of hysteresis, as indicated in figure \ref{fig:bifurc}. Finally, we note that the zero and two states are continuously connected by solutions that are unstable to equatorially asymmetric perturbations, but which can easily be computed by artificially suppressing these asymmetric flow components. Figure \ref{fig:contours} shows examples of these zero-, one- and two-vortex states, as well as one-dimensional slices of $\psi(\theta)$ at a constant $r$ roughly in the middle of the gap. The Taylor vortices show up very clearly as oscillations in these slices. The gradual linear trend that is also noticeable is due to the large-scale Ekman cells.

For $\epsilon=0.154$, the bifurcation diagram of \citep{mamun1995} details exactly the same sequence of zero-, one- and two-vortex states as our $\epsilon=0.04$ solutions. It is only subsequent bifurcations that are different; \citep{mamun1995} obtain a variety of other steady solutions, including equatorially asymmetric ones, whereas we obtain time-dependent pulse train solutions. The time-dependent solutions at $\epsilon=0.04$ are similar though to the ones at $\epsilon=0.02$, so we only studied the latter case in full detail. Similar axisymmetric time-dependent solutions were noted by \citep{yuan2004numerical} for $\epsilon=0.06$, though not discussed in any detail.

Turning then to $\epsilon=0.02$, and comparing it with $\epsilon=0.04$, the first point to note is that the one-vortex state has disappeared entirely, and the zero and two states now merge together without any unstable gap in between. That is, the two state is now effectively part of the basic state still, in the sense that no bifurcations occur as $\mathrm{Re}$ is increased up to where a two state exists. By varying $\epsilon$ between 0.02 and 0.04, it is relatively straightforward to determine where the transition from one bifurcation diagram to the other occurs. The unstable solutions connecting the zero and two states still exist at $\epsilon=0.023$, but at $\epsilon=0.022$ this unstable segment has ceased to exist, and the zero and two state merge together in a way that is not only continuous, but stable throughout the entire solution branch. Determining how far $\epsilon$ can be reduced before the one state ceases to exist is somewhat more difficult to compute, since calculations in the vicinity of turning points inevitably take very long to equilibrate. It seems though that the one state also disappears somewhere between $\epsilon=0.025$ and 0.02.  The one state disappearing, and the unstable gap between the zero and two states disappearing, are two separate phenomena though, and thus almost certainly do not happen at precisely the same critical value of $\epsilon$. However, as there is no particular interest in knowing precisely for which $\epsilon$ values either event occurs, we return instead to the single value $\epsilon=0.02$, and continue increasing $\mathrm{Re}$ until the basic state (now the two state) does eventually become unstable. This occurs at $\mathrm{Re}=15118$, where a supercritical Hopf bifurcation leads to a branch of periodic, equatorially symmetric
travelling pulse solutions, of exactly the variety explored by \citep{harris2003,bassom2004,blockley2007}. Indeed, the solution is qualitatively very similar to that in Fig.\ 3 of \citep{blockley2007}.

\begin{figure*}[ht]
 \centering
 \includegraphics[width=\textwidth]{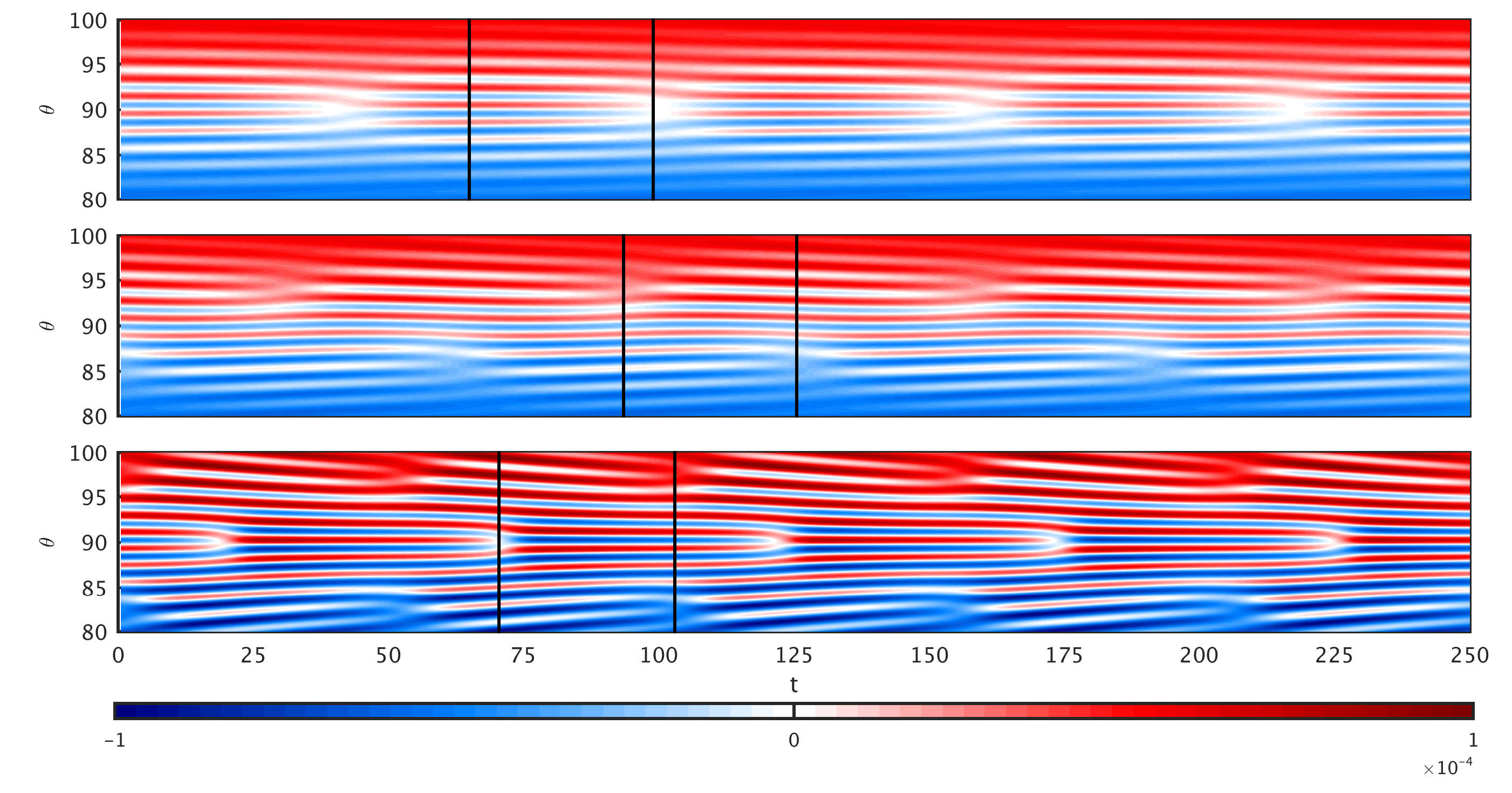}
 \caption{From top to bottom, Hovm\"oller plots of $\psi(\theta,t)$ for $\mathrm{Re}=15200$, $15250$, $15650$. The black lines denote slices of $\psi(\theta)$ used in figure
 \ref{fig:packets}.} 
 \label{fig:hov_per}
\end{figure*}

As previously seen in figure \ref{fig:contours}, all the interesting dynamics occur in $\theta$, with relatively little structure in $r$. The entire space-time dependence of the solutions can then be conveniently captured in so-called Hovm\"oller plots, showing $\psi(\theta,t)$ at fixed $r=(r_i+r_o)/2$. Figure \ref{fig:hov_per} shows results of the first few bifurcations (see again figure \ref{fig:bifurc} for the entire bifurcation diagram). The initial Hopf bifurcation from the steady two-vortex state to a time-dependent state results in Taylor vortices drifting toward the equator, coming symmetrically from both hemispheres. At the equator pairs of vortices are periodically destroyed in phase-slip events, as seen in the top row of figure \ref{fig:hov_per}.

Further increasing $\mathrm{Re}$, at $\mathrm{Re}=15245$ this equatorially symmetric solution undergoes a subcritical pitchfork bifurcation to an equatorially asymmetric perturbation. The middle row of figure \ref{fig:hov_per} shows the resulting new solution branch. We note how the previous equatorial symmetry $\psi(\theta,t)=-\psi(\pi-\theta,t)$ has been broken, but the solution still preserves the so-called shift-and-reflect symmetry $\psi(\theta,t)= -\psi(\pi-\theta,t+T/2)$, where $T$ is the period. Notice also how the phase-slips now occur several degrees off the equator, alternating in the two hemispheres. If $\mathrm{Re}$ is decreased again, this shift-and-reflect solution branch exists back down to $\mathrm{Re}=15210$, thus having a slight amount of hysteresis with the previous equatorially symmetric branch. If instead $\mathrm{Re}$ is increased, it exists up to $\mathrm{Re}=15775$. We
conjecture that the ends of the interval are both turning point bifurcations, with unstable solution branches connecting as in figure \ref{fig:bifurc}.

If $\mathrm{Re}$ is increased beyond 15775, the system switches back to an equatorially symmetric solution (again with hysteresis if $\mathrm{Re}$ is then reduced). The bottom row of figure \ref{fig:hov_per} shows these solutions. Note how phase-slips now occur alternately on the equator and $\sim7^\circ$ offset from it. Finally, as indicated in figure \ref{fig:bifurc}, the two equatorially symmetric solution branches existing for $15118\le\mathrm{Re}\le15245$ and $15575\le\mathrm{Re}\le16190$ are continuously connected by solutions that are unstable to equatorially asymmetric perturbations, but which can easily be computed by artificially suppressing these asymmetric flow components (just as before for the branch connecting the $\epsilon=0.04$ zero and two states).

\begin{figure*}[ht]
 \centering
 \includegraphics[width=\textwidth]{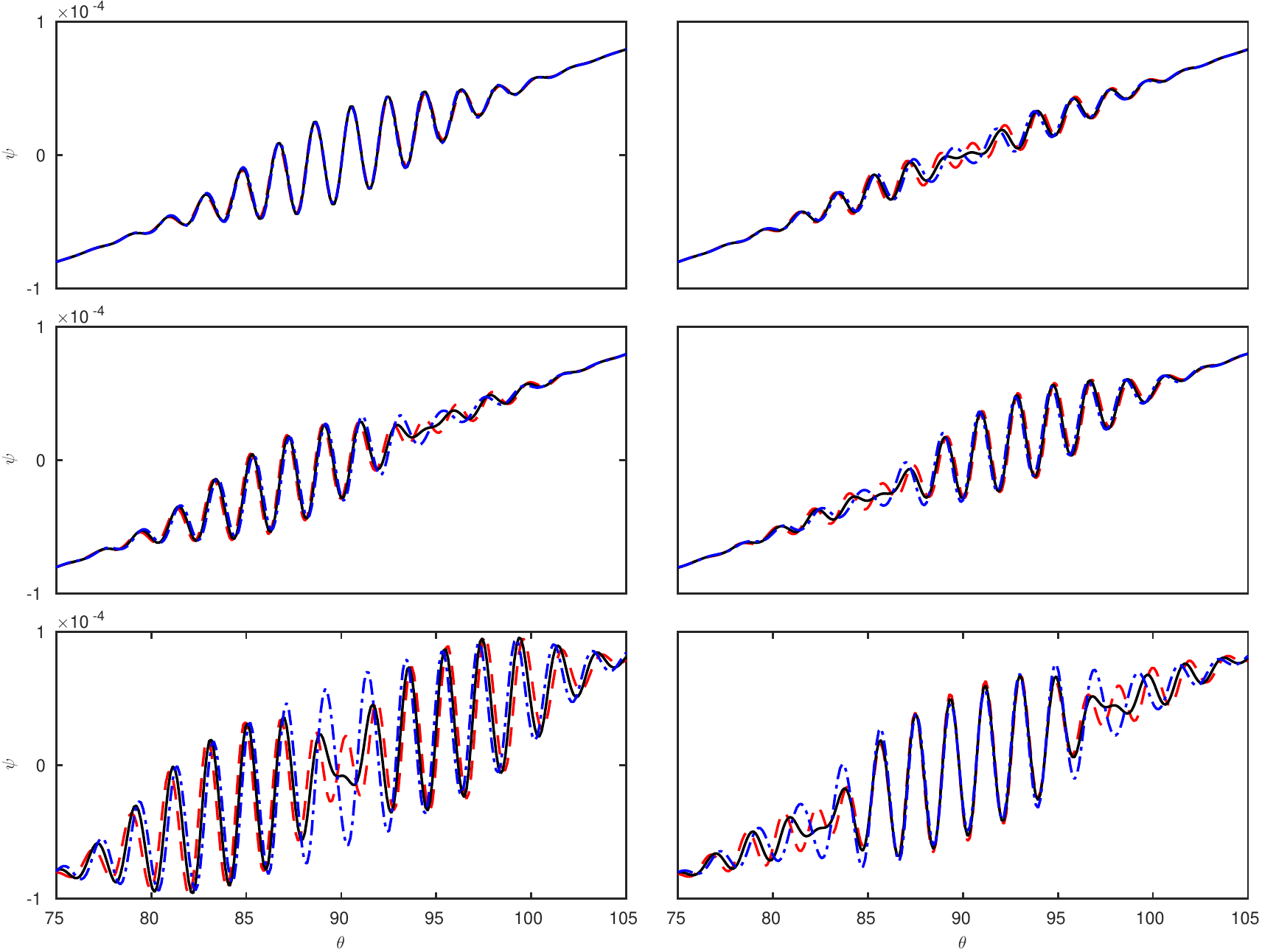}
 \caption{From top to bottom, $\mathrm{Re}=15200$, $15250$, $15650$, as in figure \ref{fig:hov_per}. The left and right panels correspond to the times indicated by the two vertical lines in figure \ref{fig:hov_per}. Each panel then shows $\psi(\theta)$ at exactly that time as a solid black line, four rotation periods before as a red dashed line, and four rotation periods after as a blue dot-dashed line.}
 \label{fig:packets}
\end{figure*}

Figure \ref{fig:packets} shows further details of these three types of solutions, and their resulting phase-slip events. The rows are the same as in figure \ref{fig:hov_per}; the left and right panels then correspond to the particular times indicated by the two black lines shown for each solution in figure \ref{fig:hov_per}. Each panel shows $\psi(\theta)$ at that time, as well as just before and after, to give a clearer indication of the nature of the time-dependence, that is, the phase-slip events. In the top row (the symmetric solution at $\mathrm{Re}=15200$), we see that between the phase-slips the pattern is essentially steady, whereas during the phase-slips the pattern right at the equator evolves very quickly. In the middle row (the shift-and-reflect solution at $\mathrm{Re}=15250$), the phase-slips again occur very quickly, in each case at the edge of the wave packet that is closer to the equator. In the bottom row (the symmetric solution at $\mathrm{Re}=15650$), the phase-slips occur alternately in the middle and the edges of the wave packet. In all cases the time-dependence is most rapid where the phase-slips are occurring, with the rest of the wave packets remaining essentially steady during the phase-slips. Finally, note how the
number of Taylor vortices within the wave packets increases as $\mathrm{Re}$ is increased. This suggests a natural explanation for the existence of the intermediate shift-and-reflect solutions; if the number of Taylor vortices naturally increases by one at a time, then there must inevitably be solutions that are not equatorially symmetric.

\begin{figure*}[ht]
 \centering
 \includegraphics[width=\textwidth]{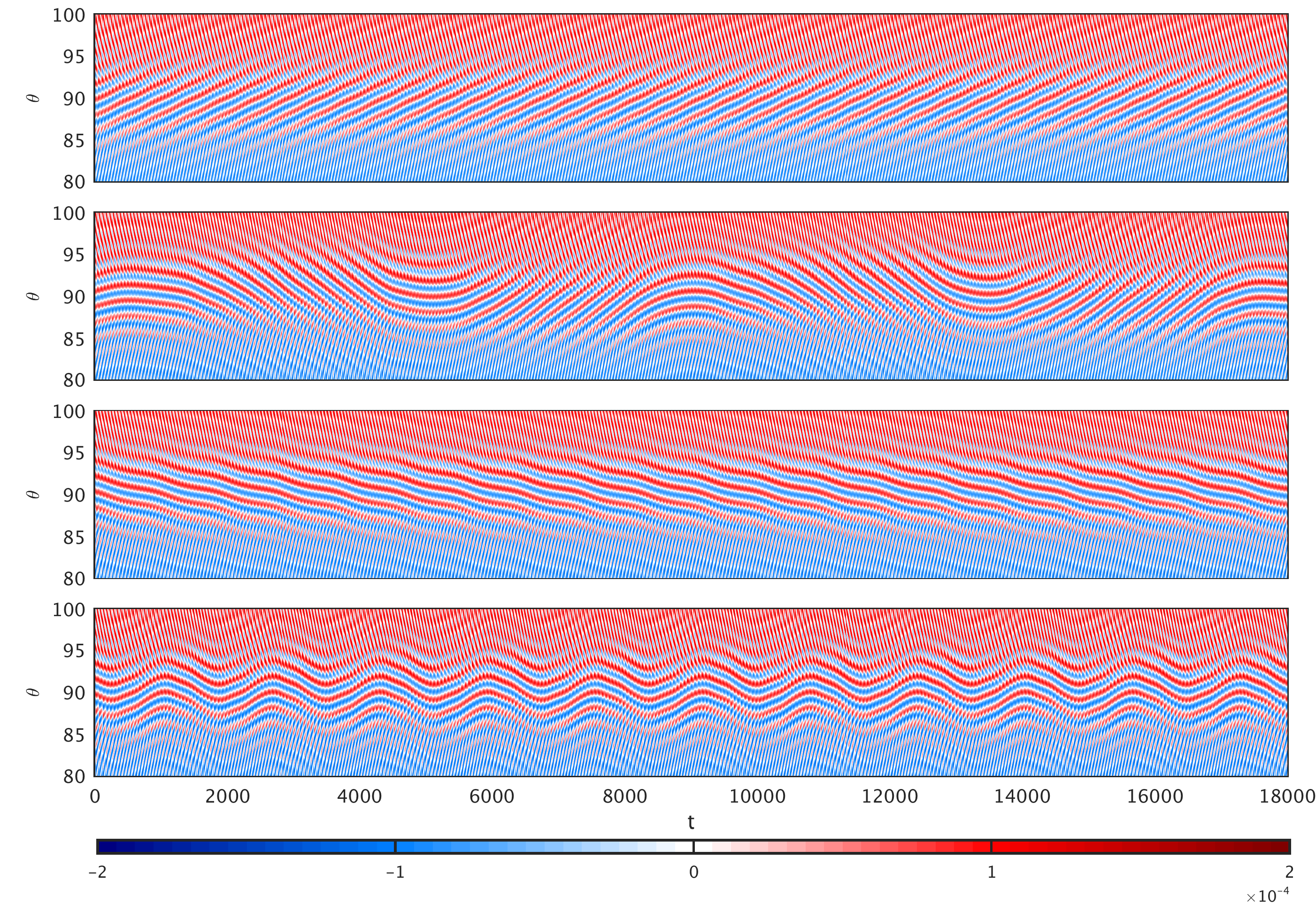}
 \caption{From top to bottom, Hovm\"oller plots of $\psi(\theta,t)$ for $\mathrm{Re}=16200$,
16250, 16275, and 16400.} 
 \label{fig:hov_quasi}
\end{figure*}

Finally, if $\mathrm{Re}$ is increased beyond 16190, these equatorially symmetric, periodic solutions become unstable to a secondary Hopf bifurcation, leading to quasi-periodic solutions. The full details of these patterns, and transitions between them, are very difficult to compute, as extremely long run times are required. We therefore did not attempt to map out the precise nature of any subsequent bifurcations in this quasi-periodic regime. Figure \ref{fig:hov_quasi} shows four examples of typical behaviour.

At $\mathrm{Re}=16200$ we still see pulse trains of Taylor vortices travelling toward the equator from both hemispheres. However, in the immediate vicinity of the equator, between approximately $85^\circ$ and $92^\circ$, there is instead a unidirectional drift, from `south' to `north'. (There is presumably also a different solution where this drift would be in the opposite direction.) The phase-slip events then occur at the boundaries, where the outer pulse trains impinge on this equatorial region. The quasi-periodicity of the solutions comes about because the drift rates of the inner and outer pulse trains do not match up in any simple ratio.

For greater $\mathrm{Re}$, the phase-slip events have disappeared entirely, and there are only pulse trains that either oscillate up and down, at $\mathrm{Re}=16250$ and 16400, or else travel unidirectionally, at $\mathrm{Re}=16275$. It is possible of course that even these unidirectionally travelling solutions would eventually reverse their direction of travel on some sufficiently long time-scale. That is, all three of these solutions could potentially be qualitatively exactly the same, just with reversal periods that vary enormously with $\mathrm{Re}$. This difficulty in knowing how long one must integrate before one has truly captured all the dynamics of the
solutions just underscores the computational difficulties in fully exploring this quasi-periodic regime.

\section{Conclusion}

The pulse train solutions we have numerically computed here are in excellent qualitative agreement with the asymptotic results of \citep{harris2000,harris2003,bassom2004,blockley2007,soward2016}. In accordance with their results, we agree also that a new `very narrow gap' regime ought to be defined, in which these pulse train solutions are the first bifurcation from the basic state. We have identified that this point occurs around $\epsilon\approx0.023$, with narrower gaps having pulse trains as the first bifurcation, but wider gaps having the first few bifurcations at least being the same as throughout the remainder of the `narrow gap' regime. Within the pulse train regime, we obtained a broad variety of solutions, including equatorially symmetric and asymmetric, as well as periodic and quasi-periodic. Finally, we note that it would be of interest to compute the possibility of non-axisymmetric solutions in some of these regimes, as observed experimentally by \citep{nakabayashi1983transition} at $\epsilon=0.08$ as well as numerically by \citep{yuan2004numerical} at $\epsilon=0.06$. Computationally though this would be even more
challenging than the calculations presented here, so any attempts in this direction are deferred to future work.

\section{Acknowledgements}

We are grateful to Nathana\"el Schaeffer for adding an F77 implementation of the Legendre transform to his SHTNS code \citep{schaeffer2013}, and general assistance in its usage. AC was supported by an STFC studentship; RH and EK were supported by STFC Grant No.\ ST/K000853/1.

\end{document}